\documentclass[aps,prl,twocolumn,superscriptaddress,showpacs,amsmath,amssymb,longbibliography]{revtex4-2}
\usepackage[utf8]{inputenc}
\usepackage{graphicx}
\usepackage{epstopdf}
\usepackage{amsmath,amsfonts,amssymb,color,graphicx,xcolor,physics}
\usepackage{hyperref}
\hypersetup
{
	colorlinks=true,
	linkcolor=blue,
	filecolor=blue,
	urlcolor=blue,
	citecolor=cyan,
}
\usepackage{dcolumn}%
\newcommand{\orcid}[1]{\href{https://orcid.org/#1}{\includegraphics[width=6.6pt]{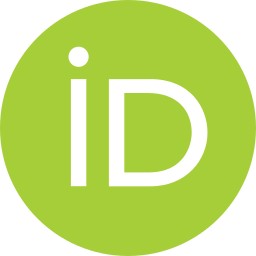}}}

\begin{document}
\title{Neural Network-Based Design of Approximate Gottesman-Kitaev-Preskill Code}
\author{Yexiong Zeng\orcid{0000-0001-9165-3995}}
\affiliation{Theoretical Quantum Physics Laboratory, Cluster for Pioneering Research, RIKEN, Wakoshi, Saitama 351-0198, Japan}
\affiliation{Quantum Computing Center, RIKEN, Wakoshi, Saitama 351-0198, Japan}

\author{Wei Qin\orcid{0000-0003-1766-8245}}%
\altaffiliation[qin.wei@tju.edu.cn]{}
\affiliation{Theoretical Quantum Physics Laboratory, Cluster for Pioneering Research, RIKEN, Wakoshi, Saitama 351-0198, Japan}%
\affiliation{Center for Joint Quantum Studies and Department of Physics, School of Science, Tianjin University, Tianjin 300350, China}
\affiliation{Tianjin Key Laboratory of Low Dimensional Materials Physics and Preparing Technology, Tianjin University, Tianjin 300350, China}

\author{Ye-Hong Chen\orcid{0000-0002-7308-2823}}
\affiliation{Fujian Key Laboratory of Quantum Information and Quantum Optics, Fuzhou University, Fuzhou 350116, China}
\affiliation{Department of Physics, Fuzhou University, Fuzhou 350116, China}
\affiliation{Theoretical Quantum Physics Laboratory, Cluster for Pioneering Research, RIKEN, Wakoshi, Saitama 351-0198, Japan}
\affiliation{Quantum Computing Center, RIKEN, Wakoshi, Saitama 351-0198, Japan}


\author{Clemens Gneiting\orcid{0000-0001-9686-9277}}
\altaffiliation[clemens.gneiting@riken.jp]{}
\affiliation{Theoretical Quantum Physics Laboratory, Cluster for Pioneering Research, RIKEN, Wakoshi, Saitama 351-0198, Japan}
\affiliation{Quantum Computing Center, RIKEN, Wakoshi, Saitama 351-0198, Japan}

\author{Franco Nori\orcid{0000-0003-3682-7432}}
\altaffiliation[fnori@riken.jp]{}
\affiliation{Theoretical Quantum Physics Laboratory, Cluster for Pioneering Research, RIKEN, Wakoshi, Saitama 351-0198, Japan}
\affiliation{Quantum Computing Center, RIKEN, Wakoshi, Saitama 351-0198, Japan}
\affiliation{Department of Physics, University of Michigan, Ann Arbor, Michigan, 48109-1040, USA}

\date{\today}
\begin{abstract} 
Gottesman-Kitaev-Preskill (GKP) encoding holds promise for continuous-variable fault-tolerant quantum computing. While an ideal GKP encoding is abstract and impractical due to its nonphysical nature, approximate versions provide viable alternatives. Conventional approximate GKP codewords are superpositions of multiple {large-amplitude} squeezed coherent states. This feature ensures correctability against single-photon loss and dephasing {at short times}, but also increases the difficulty of preparing the codewords. To minimize this trade-off, we utilize a neural network to generate optimal approximate GKP states, allowing effective error correction with just a few squeezed coherent states. We find that such optimized GKP codes outperform the best conventional ones, requiring fewer squeezed coherent states, while maintaining simple and generalized stabilizer operators. Specifically, the former outperform the latter with just \textit{one third} of the number of squeezed coherent states at a squeezing level of 9.55 dB. This optimization drastically decreases the complexity of codewords while improving error correctability.
\end{abstract}
\maketitle 

{\it Introduction}---Quantum error correction (QEC), employing syndrome measurements or environmental engineering to restore encoded quantum information, plays a pivotal role for realizing large-scale fault-tolerant quantum computing ~\cite{Chiaverini2004Dec,Schindler2011May,lidar_brun_2013,RevModPhys.87.307,Gaitan2017Jan,PhysRevLett.76.3108,PhysRevLett.85.856,PhysRevLett.105.040502,Lescanne2020May}. Notably, bosonic quantum error correction promises to enable information storage in a single bosonic mode by leveraging the infinite-dimensional Hilbert space of the mode to provide redundancy for effective error resilience ~\cite{Chuang1997Aug,PhysRevLett.131.050601,Matsuura2020Sep,Ma2020Aug,Ni2023Apr,PRXQuantum.3.020302}. The extended lifespan and the well-defined error model of superconducting microwave cavities offer practical experimental support for this type of coding~\cite{You2011Jun,RevModPhys.85.623,PhysRevB.68.064509,Ofek2016Aug,Krastanov2021Jan}.

Among the bosonic codes, the Gottesman-Kitaev-Preskill (GKP) code is distinguished by its performance in correcting arbitrary small oscillator displacement errors. For the ideal GKP code, such errors can be corrected by an appropriate QEC method, which exclusively involves Gaussian operations~\cite{Gottesman2001,Tzitrin2020,PhysRevA.108.012603,PhysRevX.6.031006,PRXQuantum.3.020334,Albert2018Mar,PhysRevLett.131.170603,Heussen2024Feb,PhysRevLett.132.130605}. The ideal GKP code is a powerful concept, yet its impracticality restricts its direct application to quantum computing. Feasible finite-energy approximate GKP states are required. The commonly used approximate GKP states are superpositions of highly squeezed coherent states, which gradually approach the ideal GKP states with increasing squeezing levels. A large squeezing, often above 9.5 dB, is required for effective QEC against single-photon loss and dephasing~\cite{Sivak2023,PRXQuantum.2.030325}.

However, raising the squeezing level increasingly disperses the approximate GKP states within the Fock state space, ultimately amplifying the effect of dephasing channels~\cite{supplement}. {For conventional GKP codes, the very small probability amplitudes  of large-amplitude squeezed coherent states critically impact the simultaneous error-correct ability for both dephasing and single-photon loss at short times.  However, these large-amplitude components are difficult to control precisely, resulting in fundamental obstacles to producing high-quality GKP codewords with superior error correction capabilities~\cite{deNeeve2022Mar,Campagne-Ibarcq2020Aug,Eickbusch2022Dec}. In particular, their optical preparation process requires breeding large-amplitude cat states, a task complicated by low success rates, limited amplitudes, and inadequate squeezing~\cite{PhysRevX.13.031001,Hastrup2022Apr,PhysRevA.97.022341,Matthew2024,Vasconcelos2010Oct}. 
}

In this letter, we use neural networks to model the coefficient functions of squeezed coherent states in approximate GKP states.  In this approach, the optimized GKP code aims to minimize the number of squeezed coherent states while maximizing error-correctability as determined by the Knill-Laflamme (KL) criterion~\cite{Knill1997Feb,PhysRevLett.84.2525}. Furthermore, we ensure that the produced approximate GKP states maintain the same distance to the ideal GKP states as the best conventional GKP code defined by a fixed coefficient function with optimum parameters. We find that GKP states optimized by the neural network outperform the error correction bound set by the best conventional GKP code while significantly reducing the number of large-amplitude squeezed coherent states. For example, at a squeezing level of $9.55$ dB, our optimized GKP states, with just \textit{seven} squeezed coherent states, surpass the best conventional GKP approximation, which requires $21$ squeezed coherent states. The optimized approximate GKP encoding also allows for simple stabilizer operators and quantum gates.
Our approach relies on the theoretical result that finite neural networks can approximate any continuous function with arbitrary precision~\cite{Hornik1991Jan,RevModPhys.91.045002,Palmieri2020Feb,Zeng2021Apr}. 

{\it Finite-Energy GKP Code}--- 
We {focus on the square GKP codewords}, defined as the common eigenstates for the operators $S_{q}=\exp(i2\sqrt{\pi}\hat{q})$ and $S_{p}=\exp(-i2\sqrt{\pi}\hat{p})$ with a shared unit eigenvalue~\cite{Shaw2024Feb}. Here, $\hat{q}$ and $\hat{p}$ are quadrature coordinates of a harmonic oscillator and satisfy the commutation relation $[\hat{q},~\hat{p}]=i$. These codewords are non-normalizable and impractical. Utilizing a superposition of squeezed coherent states, however, allows us to approximate them,	
\begin{align}
	\vert u_{L}\rangle	=\frac{1}{\mathcal{N}(u)}\sum_{k=-M}^{M}c_{k}^{(u)}\vert\alpha_{k}^{(u)},r\rangle,\,\,\,\,\,\,u\in\lbrace0,1\rbrace,
	\label{codewords}
\end{align}	
where $\vert\alpha_{k}^{(u)},r\rangle$ is a squeezed coherent state 
with squeezing magnitude $r$ (phase $\theta=0$) and displacement $\alpha_{k}^{(u)}$,  $\mathcal{N}(u)$ is the normalization coefficient, and $(2M+1)$ is the number of squeezed coherent states. Increasing the squeezing magnitude $r$ reduces the difference between the approximate and ideal GKP states. The coefficients $c^{(u)}_k$, as nonlinear functions of $\alpha^{(u)}_{k}=\sqrt{\frac{\pi}{2}}(2k+u)$ and the squeezing magnitude $r$, { are optimizable} and play a crucial role
in QEC. {Note that $\alpha^{(u)}_{k}$ is kept fixed to ensure that Eq.~(\ref{codewords}) approximates the square GKP codewords (the Wigner function is a square grid).}~The various choices for the nonlinear functions $c^{(u)}_k$ may offer alternatives that are superior to conventional GKP codewords.
For the conventional GKP code, the coefficients
$c_{k}^{(u)}$ are defined as  $c^{(u)}_k = \exp[-\pi\zeta^2(2k+u)^2/2]$\cite{PRXQuantum.2.020101,Glancy2006Jan,Fukui2017Nov}, where $\zeta^{-1}$ describes the Gaussian envelope width\cite{Tzitrin2020,Wang2022Nov}. 
Optimizing $\zeta$ then yields the best conventional GKP code for QEC performance.

The noise channel is represented as $\mathcal{N}_{t}(\hat{\rho})=\exp(\mathcal{L}t)\hat{\rho}=\sum_i \hat{A}_i(t)\hat{\rho} \hat{A}_i^{\dagger}(t)$, where $\hat{A}_i(t)$ and $\mathcal{L}$ denote the Kraus operator and Lindblad superoperator, respectively. This noise channel incorporates both single-photon loss and dephasing. In practical QEC, we focus on recovering short-term errors quickly and repeatedly.
For small time scales $\kappa\tau\ll 1$ and $\kappa_{\phi}\tau\ll 1$, we can approximate the Kraus operators as $\hat{A}_{1} = \hat{I} - \frac{\kappa\tau}{2}\hat{a}^{\dagger}\hat{a} - \frac{\kappa_{\phi}\tau}{2}(\hat{a}^{\dagger}\hat{a})^{2}$, $\hat{A}_{2} = \sqrt{\kappa\tau}\hat{a}$, and $\hat{A}_{3} = \sqrt{\kappa_{\phi}\tau}\hat{a}^{\dagger}\hat{a}$~\cite{supplement,10.21468/SciPostPhysLectNotes.70}, where $\kappa$ ($\kappa_{\phi}$) is the rate of single-photon loss (dephasing). {
With increasing the squeezing, the approximate GKP codewords exhibit a more spread-out distribution} in Fock space, indicating that dephasing becomes the dominant source of error~\cite{supplement}. Thus, finding optimal GKP codes is critical for simultaneously  correcting single-photon loss and dephasing errors while maintaining a small $M$.

\begin{figure}
	\centering
	\includegraphics[width=0.96\columnwidth]{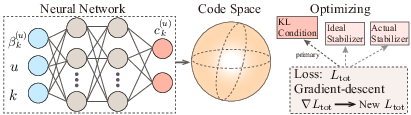}
	\vspace{-0.6\baselineskip}
	\caption{Diagram of the code optimization process. The output of the neural network contains the real and imaginary components of the coefficients $c^{(u)}_k[\beta^{(u)}_{k},u,k]$, and the corresponding input parameters are $[\beta^{(u)}_{k},u,k]$, while keeping $M$ constant. The gradient-based optimization of the loss function $L_{\text{tot}}$ determines the coefficients $c^{(u)}_k[\beta^{u}_{k},u,k]$. } 
	\label{fig1}
	\vspace{-1.2\baselineskip}
\end{figure}
\begin{figure*}
	\centering
	\includegraphics[width=2\columnwidth]{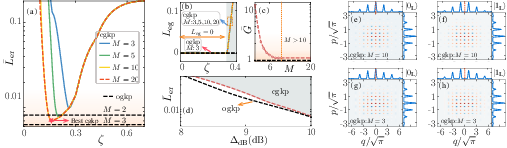}
	\vspace{-0.8\baselineskip}	
	\caption{The losses $\bar{L}_{\text{er}}$ and $L_{\text{eg}}$ for conventional GKP code versus $\zeta $ in panels (a) and (b), respectively. The black dotted lines correspond to the optimal GKP code. Note that we split here $L_{\text{tot}}$ into its components $\bar{L}_{\text{er}}$ and $L_{\text{eg}}$ to highlight their physical interpretation.
		The time scales	$\kappa\tau$ and $\kappa_{\phi}\tau$ lie within the range $[0, 0.005]$, with a squeezing strength of $\Delta_{\text{dB}}\approx 9.55$dB.
		(c) The gain $\bar{G}$, defined as the ratio associated with $L_{\text{er}}$, derived from the optimal GKP code $(M=3)$, in comparison to the most effective conventional GKP scenario given in (a). This assessment covers a larger parameter space than the training parameters (i.e., $\kappa\tau,~\kappa_{\phi}\tau\in[0,~0.01]$). 
		(d) Loss function $\bar{L}_{\text{er}}$ of the optimal GKP code obtained at $\Delta_{\text{dB}}\approx 9.55$dB against the disturbances of squeezing in the range of $8$ to $10$ dB, and compared to the best conventional GKP code. Panels (e,~f) and (g,~h) present the Wigner functions for the conventional and optimal codewords ($\vert 0_L\rangle$, $\vert1_L\rangle$) for $M=10$ and $M=3$, respectively.    
	}	
	\label{fig:schematicp2}
	\vspace{-1.2\baselineskip}
\end{figure*}

{The error-correct ability of a code can be assessed through deviations from the KL criterion~\cite{Knill1997Feb,PhysRevLett.84.2525,Terhal2020Jul}.}
Specifically, minimizing the errors $\epsilon_{ji}=\langle 1_L\vert \hat{A}^{\dagger}_j\hat{A}_i\vert 1_L\rangle-\langle 0_L\vert 
\hat{A}^{\dagger}_j\hat{A}_i\vert 0_L\rangle$ ensures equal error probabilities for the two logical basis states; $\zeta_{ji}=\langle 0_L\vert \hat{A}^{\dagger}_j\hat{A}_i\vert 1_L\rangle$ maintains the orthogonality of the error space; $\delta=\langle 0_L\vert 1_L\rangle$ keeps the logical basis orthogonality. 
If all these errors vanish (i.e., $\epsilon_{ji},~ \delta,~\zeta_{ij}=0$), {the KL condition is satisfied} and exact QEC is, in principle, possible~\cite{Leung1997Oct,PhysRevLett.104.120501,PhysRevLett.94.080501,PhysRevX.10.041018}.
However, achieving such exact QEC is challenging for approximate GKP codes at finite squeezing levels. Realistically, errors in the approximate GKP code space caused by single-photon loss and dephasing channels can only be incompletely corrected on actual experimental platforms. {Therefore, our goal is to find codewords that satisfy the KL condition to the greatest extent possible; which is equivalent to maximizing QEC performance in an error correction cycle. Consequently, we define the loss function,}		\vspace{-0.12\baselineskip}
\begin{equation}\label{klcondition}
	L_{\text{er}}=\vert\delta\vert+\sum_{ij}\left(\vert \epsilon_{ji}\vert+\vert \zeta_{ij}\vert\right).
			\vspace{-0.3\baselineskip}	\end{equation}

We evaluate $\langle u_{L}\vert \hat{S}_q\vert u_{L}\rangle$ and $\langle u_{L}\vert \hat{S}_p\vert u_{L}\rangle$ to gauge the difference between the approximate and ideal GKP code, and find
	\vspace{-0.6\baselineskip}
\begin{equation}
\langle u_{L}\vert \hat{S}_q\vert u_{L}\rangle=\exp\left(-\pi e^{-2r}\right).
\label{sqee}
	\vspace{-0.2\baselineskip}
\end{equation}
This value is solely determined by the squeezing parameter $r$ and unaffected by $M$ or the coefficients $c^{(u)}_k$. It implies that the approximation with a few squeezed coherent states can be as effective as utilizing many. {However, $\langle u_{L} \vert \hat{S}_p \vert u_{L} \rangle$ depends on $M$ and $c^{(u)}_k$, necessitating an additional cost function 
		\vspace{-0.22\baselineskip}
	\begin{equation}
L_{\text{eg}}=\sum_{u=0,1} \max(0,\exp(-\pi e^{-2r})-\langle u_{L}\vert \hat{S}_p\vert u_{L}\rangle)
	\vspace{-0.52\baselineskip}
	\end{equation}
to ensure that $\vert u_L \rangle$ are the approximate eigenstates of $\hat{S}_p$ and keep the comparability with the conventional codes. Note that the ideal stabilizers only roughly stabilize the approximate codes at finite squeezing levels. Therefore, we consider more precise stabilizer operators, 
\begin{equation}
	\begin{split}
		\hat{S}_{q,\text{ap}}=\exp[i2\sqrt{\pi}(f_{11}\hat{q}+f_{12}\hat{p})],\\ \hat{S}_{p,\text{ap}}=\exp[-i2\sqrt{\pi}(f_{21}\hat{q}+f_{22}\hat{p})], 
	\end{split}
	\label{stab}
\end{equation}
with a complex coefficient vector $\mathbf{f}=[f_{11},~f_{12},~f_{21},~f_{22}]$. We impose the condition $f_{11}f_{22}-f_{12}f_{21}=1$ to preserve the relation $\hat{S}_{q,\text{ap}}\hat{S}_{p,\text{ap}}=\hat{S}_{p,\text{ap}}\hat{S}_{q,\text{ap}}$.
We thus define the loss function \begin{equation}
	\begin{split}
		L_{\text{st}}=\sum_{u=0,1}\sum_{\hat{O}} \vert 1-\langle u_L \vert\hat{O}\vert u_L\rangle\vert^2
		\label{stabilizer_approx},
	\end{split}
\end{equation} which ensures that the approximate GKP codewords are eigenstates of Eq.~(\ref{stab}) with eigenvalue one, where \(\hat{O} \in \{\hat{S}_{q,\text{ap}},~ \hat{S}_{p,\text{ap}},~\hat{S}^{\dagger}_{q,\text{ap}}
\hat{S}_{q,\text{ap}},~ \hat{S}^{\dagger}_{p,\text{ap}}\hat{S}_{p,\text{ap}}\}\).} The resulting total loss function, then reads
{
\begin{equation}
	L_{\text{tot}} =(1 - \eta_1-\eta_2)\bar{L}_\text{er}+\eta_1L_\text{st}+ \eta_2L_{\text{eg}}, ~~~0 < \eta_{1,2} < 1,
	\label{loss_function}
\end{equation}
}where { $\bar{L}_\text{er}=\frac{1}{N}\sum_{\kappa\tau, \kappa_{\phi}\tau} L_{\text{er}}$ probes various time scales $\kappa\tau$ and $\kappa_{\phi}\tau$ to ensure that the codewords maintain a good error-correct ability over a broad time span},
and $N$ is the number of terms summed over in $\bar{L}_{\text{er}}$. Note that we possess an exact analytical expression for the loss function in Eq.~(\ref{loss_function}), avoiding the need for numerical truncation and diminishing computational cost, especially for highly squeezed codewords~\cite{supplement}.

Our protocol is illustrated schematically in Fig.~\ref{fig1}. The neural network captures the complex nonlinear function $c^{(u)}_k[\beta^{(u)}_{k},u,k]$, yielding a neural network-based quantum state, where $\beta_{k}^{(u)}=\cosh(r)\alpha^{(u)}_{k}+\sinh(r)\alpha^{(u)*}_{k}$ represents the two-photon coherent parameters for convenient computation~\cite{supplement,PhysRevA.13.2226}.  
Optimizing the neural network {and the coefficient matrix $\textbf{f}$} by minimizing the loss function in Eq.~(\ref{loss_function}) promises improved GKP codes. Note that we optimize the neural network instead of directly optimizing $c^{(u)}_k$ to achieve a fair comparison with the conventional GKP code, which maintains a specific relation between $c^{(u)}_k$ and $[\beta^{(u)}_{k},u,k]$.
Here, we use the Adam optimizer with the CosineAnnealingWarmRestarts algorithm in PyTorch to optimize the neural network and  minimize the risk of getting stuck in local minima. 
After finding the optimum codewords, we search for the optimal recovery channel $\mathcal{R}_{\text{opt}}(\cdot)$ to examine the error-correction performance of this encoding.  Maximizing the channel fidelity $F=\frac{1}{4}\sum_{ij}\vert \Tr(\hat{R}_j\hat{A}_i)\vert^2$ is a convex optimization problem with semi-definite constraints, { where $\hat{R}_j$ is the recovery operator}~\cite{PhysRevA.75.012338,Kosut2009Oct,PhysRevA.82.042321,PhysRevA.106.022431}.
{This recover channel represents the upper boundary for QEC.}
 We employ the QuTip library to solve the associated master equation~\cite{Johansson2012Aug,Johansson2013Apr,Kosut2008Jan,Taghavi2010Oct} and the Cvxpy library for the semidefinite convex optimization in Python~\cite{diamond2016cvxpy,agrawal2018rewriting}.

{\it Learning Outcomes}---
We optimize the quantum states with $r=1.1$ for example, where $r=1.1$ corresponds to the squeezing level $\approx9.5$ dB attainable in current experiments~\cite{Sivak2023,deNeeve2022Mar,Campagne-Ibarcq2020Aug}. After a meta parameters exploration, we settle for two hidden layers, each containing five neurons. {The learning rate and $(\eta_1,\eta_2)$ are $10^{-4}$ and $(0.02, 0.02)$, respectively.}
The optimized GKP code with $M=3$ \textit{(i.e., seven squeezed coherent states)} exhibits a significantly lower value of the loss function $\bar{L}_{\text{er}}$ than the conventional code with $M=10$ \textit{(i.e., 21 squeezed coherent states)}, as shown in Fig.~\ref{fig:schematicp2}(a).
The conventional code's QEC ability improves as $M$ increases, but it has an upper bound due to the constraints on the squeezing magnitude $r$ and the fixed envelope. Surprisingly, we find that the envelope exceeds this threshold and significantly decreases the number of superposed squeezed states. The optimal coefficients are listed in Tab.~\ref{table}.
{
\begin{table*}[htbp]
	\vspace{-\baselineskip}
	\centering
	\caption{Real and imaginary parts of the optimal coefficients $c^{(u)}_k[\beta^{(u)}_{k},u,k]/\mathcal{N}(u)$}
	\renewcommand{\arraystretch}{0.2}
	\begin{ruledtabular}
			\begin{tabular}{r|@{}c*{6}{@{\hspace{0pt}}c}}
				Re[$c^{(0)}$]$/\mathcal{N}(0)$ &  0.053086 & 0.22733  & 0.314502& 0.349696& 0.281129& 0.227219& 0.10026\\
				Im[$c^{(0)}$]$/\mathcal{N}(0)$ &-0.069034 & -0.219535 & -0.280702&-0.318202 &-0.254336 & -0.216339 &-0.11228  \\
				Re[$c^{(1)}$]$/\mathcal{N}(1)$ &0.124631 & 0.243408 & 0.300107& 0.278471& 0.230698& 0.16376& 0.009765  \\
				Im[$c^{(1)}$]$/\mathcal{N}(1)$ &-0.128982 &-0.226925&-0.272479 & -0.251869 &-0.200419 & -0.137301 &-0.053407
			\end{tabular}
	\end{ruledtabular}
	\label{table}
	\vspace{-1.8\baselineskip}
\end{table*}
}

The fidelity between ideal and approximate GKP states consistently surpasses the predefined threshold of $\exp(-\pi e^{-2r})$ for $M \geq 2$ [see Fig.~\ref{fig:schematicp2}(b)]. 
{It follows that} the optimal GKP code represents an approximation to the ideal GKP code comparable to the conventional GKP code while drastically reducing the number of squeezed coherent states.
The average gain $\bar{G} = \bar{L}_{\text{er}}(\text{cgkp})/\bar{L}_{\text{er}}(\text{ogkp})$, proportional to the infidelity ratio, consistently exceeds one [see Fig.~\ref{fig:schematicp2}(c)]. Hence, the optimal GKP codes are robust across a wide range of $\kappa\tau$ and $\kappa_{\phi}\tau$ values, beyond those involved in the training process.
Moreover, the optimal GKP code at $\approx9.5$ dB consistently outperforms the best conventional GKP code across a wide squeezing range of $8$dB$\sim$$10$dB, keeping it approximately optimal without additional neural network retraining [see Fig.~\ref{fig:schematicp2}(d)]. Notably, re-optimizing the neural network may yield even better results. 
Similarly, when the coefficients are constrained to the real-number domain, our conclusions still hold, albeit with diminished performance compared to the complex coefficients~\cite{supplement}.

The optimized GKP encoding minimizes unnecessary {large-amplitude} squeezed coherent state components by optimizing the envelope distribution~[see Fig.~\ref{fig:schematicp2}(e-h)]. The momentum marginals of the Wigner functions are invariant, consistent with the description of Eq.~(\ref{sqee}). However, for the conventional GKP code, substantial squeezed coherent state components are essential to correct single-photon loss and dephasing, even if these are small disturbances.  In particular, minor coefficient perturbations can substantially deteriorate the QEC ability in the conventional GKP code, while the optimized GKP code is more robust~\cite{supplement}.   {The optimal codewords significantly mitigate the challenge of preparing the encoded states by substantially reducing the need for generating large-amplitude cat states in optical systems and the dependence on precise control in superconducting systems~\cite{deNeeve2022Mar,Campagne-Ibarcq2020Aug,Eickbusch2022Dec,PhysRevX.13.031001,Hastrup2022Apr,PhysRevA.97.022341,Matthew2024,Vasconcelos2010Oct}.}

 
{The operators in Eq.~(\ref{stab}) can approximately stabilize the codewords, as indicated by the loss function $L_{\text{st}} \approx 1.6 \times 10^{-3}$, with each term in Eq.~(\ref{stabilizer_approx}) reaching $10^{-4}$. The corresponding coefficient matrix is $f = \left[ \begin{matrix} 1.000214 + 0.000054i & -0.000001 + 0.110828i \\ 0.002603 - 0.025265i & 1.002585 + 0.00023i \end{matrix} \right]$, which describes} a small deviation from the ideal stabilizer operators. {Additionally, the values $\langle u \vert \hat{\sigma}_z \vert u \rangle \approx (-1)^u 0.99$, $\langle u \vert \hat{\sigma}_x \vert v \rangle \approx 0.99$ ($u \neq v$), and $\parallel \hat{\sigma}_{x/z} \vert u \rangle \parallel \approx 1$ suggest that the Pauli operators are given by $\hat{\sigma}_z = \hat{S}^{1/2}_{q,\text{ap}}$ and $\hat{\sigma}_x = \hat{S}^{1/2}_{p,\text{ap}}$.}
{We can thus use the operators $\hat{d}_1 = \ln(\hat{S}_{q,\text{aq}})$ and $\hat{d}_2 = \ln(\hat{S}_{q,\text{aq}})$ coupled with an auxiliary qubit to effectively design the stabilizer protocols, where $\hat{d}_j \vert u_L \rangle \approx 0$. The Hamiltonian is \(\hat{H} = \hat{d}_j \hat{b}^{\dagger}_t + \hat{d}^{\dagger}_j \hat{b}_t\), where \(\hat{b}_t\) describes a highly dissipative auxiliary system. Applying the Trotter decomposition to the unitary operator \(\hat{U} = \mathcal{T} \exp\left(-i \int_0^t \hat{H}(\tau) d\tau\right)\), we obtain the Big-Small-Big and Small-Big-Small protocols~\cite{PhysRevLett.125.260509}.} Optimized codes with other squeezing levels or real coefficients share the same stabilizer operators in Eq.~(\ref{stab}), with the sole difference being the coefficient matrix $\textbf{f}$~\cite{supplement}.
%

\begin{figure}
	\centering
	\includegraphics[width=0.92\columnwidth]{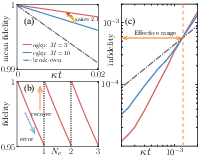}
	\vspace{-0.8\baselineskip}	
	\caption{(a) {Mean fidelity evolution for 50 error correction with $\kappa\tau=0.0004$ and $\kappa_{\phi}\tau=0.0004/1.5$}. The gain, defined as the infidelity ratio between the conventional and optimal GKP code, indicates the recovery operation efficiency achieved through convex problem-solving. (b) Fidelity fluctuations during the error and recovery processes for the initial state $(\vert 0_\text{L}\rangle+\vert 1_\text{L}\rangle)/\sqrt{2}$ within the code space. This offers a detailed portrayal of the recovery processes for the optimal GKP code as shown in (a). {(c) Comparison of the wider time scales for the QEC boundary between the optimal and conventional codewords in a single QEC cycle with $\kappa/\kappa_{\phi}=1.5$.}  		
	}	
	\label{fig:schematicp3}
	\vspace{-1.2\baselineskip}
\end{figure}
{\it Quantum Error Correction Across Multiple Cycles}--- 
{A single QEC cycle can only protect information over a short duration; multiple cycles are required to uphold the encoded information for a long time. }
We evaluate the optimized GKP encoding and the best conventional GKP code for a multiple error correction process. The entire QEC procedure can be expressed as 
\begin{equation}
	\mathcal{E}^{N_c}(\hat{\rho})=(\mathcal{R}\circ \mathcal{N})^{N_c}(\hat{\rho})
\end{equation}
where $N_c$ represents the number of QEC cycles.  We obtain the optimal recovery channel ({which puts an upper bound} on
the achievable fidelity) by solving a semi-definite convex optimization problem~\cite{Cafaro2014Feb,PhysRevA.65.030302,PhysRevLett.131.050601,supplement}. Additionally, the dephasing rate is typically lower than the single-photon loss, as demonstrated in experiments with $\kappa/\kappa_{\phi}\approx1.5$~\cite{Sivak2023}. Hence, we choose this ratio to determine the mean fidelity versus $N_{c}$ under the optimal recovery channel. The mean fidelity is estimated using the six-point intersection of the coherent Bloch sphere face and axis
~\cite{PhysRevLett.131.050601}. 

As shown in Fig.~\ref{fig:schematicp3}(a), the optimized GKP encoding has a {gain of $\approx2.1$  compared to the best conventional GKP code for a reasonable time scale.} In Fig.~\ref{fig:schematicp3}(b), we depict how the fidelity evolves from a specific initial state throughout three error correction cycles. Specifically, the encoded state evolves freely over a short period of time, resulting in errors and a slow fidelity decrease; after a specific time interval, errors are detected, followed by a recovery procedure that restores the fidelity to a value near one. This error correction cycle is conducted iteratively to ensure long-term data storage.  
{ Figure~\ref{fig:schematicp3}(c) demonstrates that the optimal codewords achieve a higher upper bound compared to the conventional ones, enabling greater error tolerance in imperfect recovery processes across various time scales within the effective range, where the performance exceeds the break-even point (at which logical qubits begin to outperform physical qubits).  
}

{\it Discussion}---We used a neural network to find the optimal GKP code when the encoded system suffers single-photon loss and dephasing. Our results show that the optimized GKP encoding requires just one-third of the number of squeezed coherent states of the best conventional GKP code to achieve better QEC ability {and retain the general and simple stabilizer operators. These squeezed coherent states are arranged in close vicinity to the squeezed vacuum state, eliminating the need for numerous large-amplitude squeezed coherent states. Consequently, the optimized codewords substantially reduce the challenges of the state preparation, offering a superior alternative to conventional GKP codes. Additionally, our method can be adapted to other types of GKP codes, such as rectangular and hexagonal GKP codes, and it can serve as a reference for future corrections of single-photon loss and dephasing, as well as for developing new codes with simplified gate operations.}  In conclusion, our technique may significantly reduce the threshold for continuous-variable error correction.


\begin{acknowledgments}
F.N. is supported in part by: 
Nippon Telegraph and Telephone Corporation (NTT) Research, 
the Japan Science and Technology Agency (JST) 
[via the CREST Quantum Frontiers program Grant No. JPMJCR24I2, 
the Quantum Leap Flagship Program (Q-LEAP), and the Moonshot R\&D Grant Number JPMJMS2061], 
and the Office of Naval Research (ONR) Global (via Grant No. N62909-23-1-2074). 
 C.G. is partly supported by a RIKEN Incentive Research Project Grant. W.Q. acknowledges support of the National Natural Science Foundation of China (NSFC) (via Grants
No.~0401260012 and No.~62131002). Y.-H.C. was supported by the National Natural Science Foundation of China under
Grant No.~12304390. 
\end{acknowledgments}


%


\end{document}


\title{Supplemental Information for ``Neural Network-Based Design of Optimal Approximate GKP Codes``}

\author{Yexiong Zeng\orcid{0000-0001-9165-3995}}
\affiliation{Theoretical Quantum Physics Laboratory, Cluster for Pioneering Research, RIKEN, Wakoshi, Saitama 351-0198, Japan}
\affiliation{Quantum Computing Center, RIKEN, Wakoshi, Saitama 351-0198, Japan}

\author{Wei Qin\orcid{0000-0003-1766-8245}}%
\altaffiliation[qin.wei@tju.edu.cn]{}
\affiliation{Theoretical Quantum Physics Laboratory, Cluster for Pioneering Research, RIKEN, Wakoshi, Saitama 351-0198, Japan}%
\affiliation{Center for Joint Quantum Studies and Department of Physics, School of Science, Tianjin University, Tianjin 300350, China}
\affiliation{Tianjin Key Laboratory of Low Dimensional Materials Physics and Preparing Technology, Tianjin University, Tianjin 300350, China}

\author{Ye-Hong Chen\orcid{0000-0002-7308-2823}}
\affiliation{Fujian Key Laboratory of Quantum Information and Quantum Optics, Fuzhou University, Fuzhou 350116, China}
\affiliation{Department of Physics, Fuzhou University, Fuzhou 350116, China}
\affiliation{Theoretical Quantum Physics Laboratory, Cluster for Pioneering Research, RIKEN, Wakoshi, Saitama 351-0198, Japan}
\affiliation{Quantum Computing Center, RIKEN, Wakoshi, Saitama 351-0198, Japan}

\author{Clemens Gneiting\orcid{0000-0001-9686-9277}}
\altaffiliation[clemens.gneiting@riken.jp]{}
\affiliation{Theoretical Quantum Physics Laboratory, Cluster for Pioneering Research, RIKEN, Wakoshi, Saitama 351-0198, Japan}
\affiliation{Quantum Computing Center, RIKEN, Wakoshi, Saitama 351-0198, Japan}

\author{Franco Nori\orcid{0000-0003-3682-7432}}
\altaffiliation[fnori@riken.jp]{}
\affiliation{Theoretical Quantum Physics Laboratory, Cluster for Pioneering Research, RIKEN, Wakoshi, Saitama 351-0198, Japan}
\affiliation{Quantum Computing Center, RIKEN, Wakoshi, Saitama 351-0198, Japan}
\affiliation{Department of Physics, University of Michigan, Ann Arbor, Michigan, 48109-1040, USA}

\maketitle
\tableofcontents

\section{Defining the approximate square GKP codewords}\label{sec1}
Here, we give the form of the approximate GKP codewords and analyze their features. Specifically, the approximate GKP state can be expanded as a superposition of  squeezed coherent states~\cite{PhysRevA.64.012310}
\begin{equation}
	\left\vert u_{L}\right\rangle	=\frac{1}{\mathcal{N}(u)}\sum_{k=-M}^{M}c_{k}^{(u)}\left\vert\alpha_{k}^{(u)},r\right\rangle, \,\,\,\,\, u\in\left\lbrace0,~1\right\rbrace,
	\label{codewords}
\end{equation}
where $\left\vert\alpha_{k}^{(u)},r\right\rangle=\hat{D}(\alpha_{k}^{(u)})\hat{S}\left(r\right)\left\vert0\right\rangle$ is a squeezed coherent state with { the displacement $\hat{D}(\alpha)=\exp\left(\alpha \hat{a}^{\dagger} -\alpha^{*}\hat{a}\right)$ and squeezing $\hat{S}(r)=\exp\left[\frac{1}{2}\left(r^{*}\hat{a}^{2}-r \hat{a}^{\dagger2}\right)\right]$ operators}, $\mathcal{N}(u)$ is the coefficient for normalization, and the coefficients $c^{(u)}_k$ are functions of the parameters $[\beta_{k}^{(u)}, u, k]$~[i.e.,~$c^{(u)}_k=f(\beta_{k}^{(u)},~k,~u)$] with the two-photon coherent parameter $\beta_{k}^{(u)}=\cosh(r)\alpha_{k}^{(u)}+\sinh(r)\alpha_{k}^{(u){*}}$~\cite{PhysRevA.13.2226}.  {
	Specifically, we rewrite the squeezed coherent state as \(\vert \alpha^{(u)}_k, r \rangle = \hat{S}(r)\hat{S}(-r)\hat{D}(\alpha_{k}^{(u)})\hat{S}(r)\vert 0 \rangle = \hat{S}(r)\hat{D}(\beta^{(u)}_k)\vert 0 \rangle\), which is the eigenstate of the squeezed mode \(\hat{b} = \hat{S}(r)\hat{a}\hat{S}(-r)\) with eigenvalue \(\beta^{(u)}_k\). This allows us to calculate analytical solutions for the expectation values of operators in the squeezing frame and avoids numerical truncations, that is, \(\alpha_{k}^{(u)} \rightarrow \beta_{k}^{(u)}\) and \(\hat{a} \rightarrow \hat{b}\).} 
Here, we focus on the square GKP codewords and assume a real squeezing parameter $ r $.  According to  Eq.~(\ref{codewords}), we can express the normalization coefficient $\mathcal{N}(u)$ as: 
\begin{equation}
	\begin{split}
\mathcal{N}^{2}(u)	=&\sum_{k,l=0}f^{*}\left(\beta_{k}^{(u)},k,u\right)f\left(\beta_{l}^{(u)},l,u\right)\left\langle\alpha_{k}^{(u)},r\large\vert\alpha_{l}^{(u)},r\right\rangle
\\
=	&\sum_{k,l=0}f^{*}\left(\beta_{k}^{(u)},k,u\right)f\left(\beta_{l}^{(u)},l,u\right)\exp[-\frac{\left\vert\beta_{k}^{(u)}\right\vert^{2}+\left\vert\beta_{l}^{(u)}\right\vert^{2}}{2}+\beta^{(u)*}_{k}\beta_{l}^{(u)}].
	\end{split}
\label{normal_coef}
\end{equation}
The codewords in Eq.~(\ref{codewords}) can be regarded as the shared approximate eigenstates of the two stabilizer operators of the square GKP code  
\begin{equation}
	\hat{S}_{q}=\exp\left(i2\sqrt{\pi}\hat{q}\right)=\hat{D}\left(\sqrt{2\pi }i\right),\,\,\,\,\,\,\,\,\,\,\,\,\,\,\ \hat{S}_{p}=\exp\left(-i2\sqrt{\pi}\hat{p}\right)=\hat{D}\left(\sqrt{2\pi}\right),
\end{equation}
with the same eigenvalue one, where $\hat{q}=(\hat{a}^{\dagger}+\hat{a})/\sqrt{2}$ and $\hat{p}=(\hat{a}-\hat{a}^{\dagger})/\sqrt{2}i$ represent the position and momentum operators, respectively.
Using the relation  $\hat{D}(\beta)\hat{D}(\alpha)=\exp\left[(\beta\alpha^*-\beta^*\alpha)/2\right]\hat{D}(\alpha+\beta)$, we can derive the following expressions:
\begin{equation}
	\begin{split}
		\hat{S}_q\left\vert\alpha_{l}^{(u)},r\right\rangle&=\hat{D}\left(\sqrt{2\pi }i\right)\hat{D}\left(\alpha_{l}^{(u)}\right)\hat{S}\left(r\right)\left\vert0\right\rangle=\exp\left[\frac{\sqrt{2\pi}i}{2}\left(\alpha^{(u)*}_l+\alpha^{(u)}_l\right)\right]\left\vert\alpha_{l}^{(u)}+\sqrt{2\pi}i,r\right\rangle,	\\
		\hat{S}_p\left\vert\alpha_{l}^{(u)},r\right\rangle&=\hat{D}\left(\sqrt{2\pi }\right)\hat{D}\left(\alpha_{l}^{(u)}\right)S\left(r\right)\left\vert0\right\rangle=\exp\left[\frac{\sqrt{2\pi}}{2}\left(\alpha^{(u)*}_l-\alpha^{(u)}_l\right)\right]\left\vert\alpha_{l}^{(u)}+\sqrt{2\pi},r\right\rangle.
	\end{split}
\label{action_state}
\end{equation}
By combining the Eqs.~(\ref{codewords}) and (\ref{action_state}), we obtain the inner products 
\begin{equation}
	\begin{split}
		\left\langle u_{L}\left\vert \hat{S}_q\right\vert u_{L}\right\rangle
		=&\exp\left(-\pi e^{-2r}\right),
\\
		\left\langle u_{L}\left\vert \hat{S}_p\right\vert u_{L}\right\rangle
		=&\frac{1}{\mathcal{N}^{2}(u)}\sum_{k,l=0}f^{*}\left(\beta_{k}^{(u)},k,u\right)f\left(\beta_{l }^{(u)},l,u\right)\exp[-\frac{\left\vert\beta_{k}^{(u)}\right\vert^{2}+\left\vert\beta_{l}^{(u)'}\right\vert^{2}}{2}+\beta^{(u)*}_{k}\beta_{l}^{(u)'}],
	\end{split}
\label{sq_and_se}
\end{equation}
where we have defined the relation $ \beta_{l}^{(u)'}= \beta_{l}^{(u)}+\sqrt{2\pi}e^r$ and $\left(\alpha^{(u)}_{k}\sqrt{\frac{2}{\pi}}-u\right)/2\in \mathbb{Z}$. {Note that we focus on the most general form of the GKP code, more specifically the \textit{square} GKP code, whose Wigner function is a square grid. The positions of the grid points in this code are determined by the parameters \(\alpha^{(u)}_k\). Thus, we have fixed these parameters to ensure a precise approximation and to maintain the consistency with the ideal square GKP code.} The difference between the approximate GKP codewords and the eigenstates of the operator $\hat{S}_q$ [i.e., the right side of the first line in Eq.~(\ref{sq_and_se})] depends solely on the squeezing parameter $r$. However, the gap between the approximate GKP codewords and the eigenstate of the operator $\hat{S}_p$ is determined by the squeezing strength $r$ and the coefficients $c^{(u)}_k$ of the squeezed coherent states. These results provide us with the potential to obtain an optimal code. {
	Since the distance between the \textit{ideal} GKP code and the \textit{approximate} GKP code is determined by the squeezing strength, and  the level of approximation along the $\hat{S}_q$ direction is fixed, we only need to ensure that $\hat{S}_p$ meets or exceeds this level of approximation to qualify as a square GKP code.
	Therefore, we strive to keep $\langle u_{L} \vert \hat{S}_p \vert u_{L} \rangle\geqslant \exp(-\pi e^{-2r})$, thereby ensuring a good approximation to the ideal codewords. To achieve this, we incorporate the following component into the loss function:
\begin{equation}\label{SQE}
L_{\text{eg}}=\sum_{u=0,1} \max\left(0,\exp(-\pi e^{-2r})-\left\langle u_{L}\left\vert \hat{S}_p\right\vert u_{L}\right\rangle\right)
\end{equation}

}
\section{Quantum error correction ability}
{
	\begin{figure}[hpbt]
		\includegraphics[width=1\columnwidth]{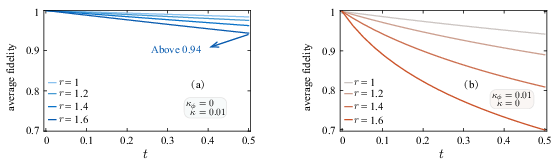}
		\caption{{
        The average fidelity \(\bar{F}\) of the logical code space versus time $t$ for various squeezing strengths: (a) \(\kappa_{\phi}=0\) and \(\kappa=0.01\); (b) \(\kappa_{\phi}=0.01\) and \(\kappa=0\). 
		\label{dispersive_foce_space}}}
	\end{figure}
}
This section provides a detailed calculation of the QEC ability. The primary source of errors in a bosonic mode are single-photon loss and dephasing. Hence, the system dynamics is governed by the Lindblad master equation
\begin{equation}
		\frac{\text{d\ensuremath{\hat{\rho}}}}{\text{dt}}=\frac{\kappa}{2}\mathcal{D}\left[\hat{a}\right]+\frac{\kappa_{\phi}}{2}\mathcal{D}\left[\hat{a}^{\dagger}\hat{a}\right], \,\,\,\,\,\,
		\mathcal{D}\left[\hat{x}\right]=2\hat{x}\hat{\rho} \hat{x}^{\dagger}-\hat{x}^{\dagger}\hat{x}\hat{\rho}-\hat{\rho} \hat{x}^{\dagger}\hat{x}.
\label{master_equation}
\end{equation}
{
To demonstrate the effects of dephasing and single-photon loss, we evaluate the evolution of the average fidelity for the conventional GKP codespace, as shown in Fig.~\ref{dispersive_foce_space}. The  average fidelity is   \(\bar{F} = \frac{1}{6} \sum_i \Tr[\hat{\rho}_i \mathcal{N}(\hat{\rho}_i)]\), where \(\mathcal{N}(\cdot)\) represents the dynamical map governed by the master equation in Eq.~(\ref{master_equation}) and \(\hat{\rho}_i\) denotes the six Pauli eigenstates of the logical space. The impact of the dephasing channel becomes more pronounced with increasing the degree of squeezing. To elucidate the effect of dephasing in detail, we provide an analytical solution for the dephasing channel in the Fock space
\begin{equation}
	\mathcal{N}_{\phi}(\rho)=\sum_{nm}\exp{\left[-\frac{\kappa_{\phi}}{2}(n-m)^2\right]}\langle n\vert\rho\vert m\rangle\vert n\rangle\langle m\vert,
\end{equation}
where the dynamic map $\mathcal{N}_{\phi}(\cdot)$ describes the pure dephasing process. As the squeezing degree increases, the distribution of codewords in the Fock space becomes more dispersed (i.e., the magnitude of \(\vert n-m \vert\) increases), which exacerbates the detrimental effects of dephasing.
}

By considering short times $\kappa\tau,~\kappa_{\phi}\tau\ll1$, we can expand the time-dependent density matrix $\rho\left(\tau\right)$ by using the Kraus operators $\hat{A}_k(\tau)$~\cite{10.21468/SciPostPhysLectNotes.70} 
\begin{equation}
		\begin{split}
		\hat{\rho}(\tau)=&\sum_{k}\hat{A}_{k}(\tau)\hat{\rho}(0)\hat{A}_{k}(\tau)^{\dagger}+\mathcal{O}(\tau^2),
		\\
		\hat{A}_{1}(\tau)=\sqrt{\hat{I}-\kappa\tau \hat{a}^{\dagger}\hat{a}-\kappa_{\phi}\tau(\hat{a}^{\dagger}\hat{a})^{2}}\approx \hat{I}-&\frac{\kappa\tau}{2}\hat{a}^{\dagger}\hat{a}-\frac{\kappa_{\phi}\tau}{2}(\hat{a}^{\dagger}\hat{a})^{2},\,\,\,\,\,\,\,\,\,\,\,\,\,
		\hat{A}_{2}(\tau)=\sqrt{\kappa\tau}\hat{a},\,\,\,\,\,\,\,\,\,\,\,\,\,  \hat{A}_{3}(\tau)=\sqrt{\kappa_{\phi}\tau}\hat{a}^{\dagger}\hat{a}.
	\end{split}
\label{master_kraus_operator}
\end{equation}
We consider an approximate QEC characterized by a finite squeezing amplitude $r$ and the form of the codewords; that is, the codewords satisfies approximately the Knill-Laflamme condition
	\begin{equation}\label{KLcondition}
	\hat{P}_C\hat{A}^{\dagger}_i\hat{A}_j\hat{P}_C=T_{ij}\hat{P}_c+\hat{\Delta}_{ij},
	\end{equation}
where $\hat{P}_C$ is the projector onto the code space, $T_{ij}$ are the elements of a Hermitian matrix $T$, and $\hat{\Delta}_{ij}$ is a residual error operator.
We use the eigendecomposition of the $T$ matrix, $T=U\Lambda U^{\dagger}$, and then remodel the Kraus operators as $\hat{F}_i=\sum_kU_{ki}\hat{A}_{k}$,  resulting in
\begin{equation}\label{key}
\sum_{i}\hat{F}_i\hat{\rho} \hat{F}^{\dagger}_i=\sum_{kl}\sum_iU_{ki}U^*_{li}\hat{A}_k\hat{\rho} \hat{A}^{\dagger}_l=\sum_{k}\hat{A}_k\hat{\rho} \hat{A}^{\dagger}_k.	
\end{equation}
With this, we can further expand the Knill-Laflamme condition as~\cite{PhysRevLett.84.2525}
\begin{equation}\label{key}
\hat{P}_C\hat{F}^{\dagger}_i\hat{F}_j\hat{P}_C=\sum_{kl}U^*_{ki}U_{lj}\hat{P}_C\hat{A}^{\dagger}_k\hat{A}_l\hat{P}_C=\sum_{kl}U^*_{ki}U_{lj}(T_{kl}\hat{P}_C+\hat{\Delta}_{kl})=\delta_{ij}\Lambda_{jj}\hat{P}_C+\hat{\tilde{\Delta}}_{ij},
\end{equation}
where we have defined $\hat{\tilde{\Delta}}_{ij}=\sum_{ki}U^*_{ki}U_{lj}\hat{\Delta}_{kl}$. 

We can employ the recovery operator $\hat{R}_{j}$ to correct the error resulting from $\hat{A}_{i}(\tau)$. The recovery operator is not unique. If we consider the recovery operator $\hat{R}_i=\hat{P}_C\hat{F}^{\dagger}_i/\sqrt{\Lambda_{ii}}$ and $\hat{R}_0=\sqrt{\hat{I}-\sum_i\hat{R}^{\dagger}_i\hat{R}_i}$ to correct the errors, we obtain the recovery channel
 \begin{equation}\label{key}
 	\begin{split}
 		\mathcal{E}(\hat{\rho})=\mathcal{R}\circ\mathcal{N}(\hat{\rho})=\sum_{ij}\frac{1}{\Lambda_{ii}}\hat{P}\hat{F}^{\dagger}_i\hat{F}_j\hat{\rho} \hat{F}^{\dagger}_j\hat{F}_i\hat{P}=	\sum_{i}\Lambda_{ii}\hat{\rho}+\sum_{ij}\hat{\tilde{\Delta}}_{ij}\hat{\rho}+\hat{\rho}\hat{\tilde{\Delta}}_{ji}+\frac{1}{\Lambda_{jj}}\hat{\tilde{\Delta}}_{ij}\hat{\tilde{\Delta}}_{ji}+\hat{R}_0\mathcal{N}(\hat{\rho})\hat{R}^{\dagger}_0, \\
 		\end{split}
  \end{equation}
which indicates that if the Knill-Laflamme criteria is satisfied (i.e.,  $\hat{\Delta}_{ij}=\hat{\tilde{\Delta}}_{ij}\rightarrow0$), the information can be corrected fully (i.e., $\mathcal{R}(\hat{\rho})=\mathcal{N}(\hat{\rho})=\hat{\rho}$).

Hence, we aim to bring the approximate GKP codewords as close as possible to the Knill-Laflamme condition
\begin{equation}
	\epsilon_{ji}=\langle1_{L}\vert \hat{A}_{j}^{\dagger}\hat{A}_{i}\vert1_{L}\rangle-\langle0_{L}\vert \hat{A}_{j}^{\dagger}\hat{A}_{i}\vert0_{L}\rangle, \,\, \zeta_{ji}=\langle 0_L\vert \hat{A}_{j}^{\dagger}\hat{A}_{i}\vert 1_L\rangle, 
	\,\, \zeta^*_{ij}=\langle 1_L\vert \hat{A}_{j}^{\dagger}\hat{A}_{i}\vert 0_L\rangle,\,\,\delta=\langle 0_L\vert 1_L\rangle.
	\label{kl_condition}
\end{equation}
It should be noted that the above recovery may not be the optimal solution.
 We can address this by solving convex optimization problems to obtain the optimal channel fidelity~\cite{PhysRevA.65.030302,PhysRevA.75.012338,Kosut2009Oct}.
 To ensure that the logical basis vectors meet the Knill-Laflamme criteria as close as possible, we define the following loss function
 \begin{equation}
 	\bar{L}_{\text{er}} =\frac{1}{N}\sum_{\kappa\tau, \kappa_{\phi}\tau} L_{\text{er}}
 	\label{loss_function},
 \end{equation}
where we have considered $L_{\text{er}}$ as
 \begin{equation}\label{klcondition}
	L_{\text{er}}=\vert\delta\vert+\sum_{ij}\left(\vert \epsilon_{ji}\vert+\vert \zeta_{ij}\vert\right).	
\end{equation}
Here $N$ is the number of terms summed over in $\bar{L}_{\text{er}}$ to guarantee the loss function can be sufficiently minimal. { Moreover, we calculate the following relation
	\begin{equation}
		\hat{P}_C\hat{A}^{\dagger}_j\hat{A}_i\hat{P}_C=\langle 0_L\vert\hat{A}^{\dagger}_j\hat{A}_i\vert 0_L\rangle\hat{P}_C+\epsilon_{ji}\vert 1_L\rangle\langle 1_L\vert+\zeta^*_{ij}\vert 1_L\rangle\langle 0_L\vert+\zeta_{ji}\vert 0_L\rangle\langle 1_L\vert,
		\label{Kl_conditon}
	\end{equation}
which implies that $\epsilon_{ji}$ and $\zeta_{ji}$ describe the Pauli $\hat{\sigma}_z$ errors and also $\hat{\sigma}_x$ or $\hat{\sigma}_y$ errors,~ respectively. If  $\zeta_{ji}$ and $\epsilon_{ji}$ vanish, the KL conditions are satisfied (i.e., no noise bias).
These noise biases are reduced roughly equally during the optimization process due to the consistent weights of errors in the loss function, ensuring that the final noise bias reaches small values.}
Moreover, the Eq.~(\ref{kl_condition}) can be expressed as the sum of $\langle u_L\vert \hat{K}^{\dagger}_i\hat{K}_j\vert v_L\rangle$, where $\hat{K}_i$ and $\hat{K}_j$ are operators selected from the set $\lbrace{\hat{I}, \, \hat{a}, \, \hat{a}^{\dagger}\hat{a}, \, \hat{a}^{\dagger}\hat{a}\hat{a}^{\dagger}\hat{a}}\rbrace$ and $u$,~$v$ are binary values (i.e., $u,v\in\lbrace 0,~1\rbrace$). Since numerical truncation can be computationally expensive and memory-intensive, we perform an analytical calculation of the parameter $\langle u_L\vert \hat{K}^{\dagger}_i\hat{K}_j\vert v_L\rangle$ to simplify the numerical simulation. We can expand $\langle u_L\vert \hat{K}^{\dagger}_i\hat{K}_j\vert v_L\rangle$ into the form
\begin{equation}
	\begin{split}
		\left\langle u_L\left\vert \hat{K}^{\dagger}_i\hat{K}_j\right\vert v_L\right\rangle=&\frac{1}{\mathcal{N}(u)\mathcal{N}(v)}\sum_{k,l=0}f^*\left(\alpha^{(u)}_k,r\right)f\left(\alpha^{(v)}_l,r\right)\left\langle\alpha^{(u)}_k,r\left\vert \hat{K}^{\dagger}_i\hat{K}_j \right\vert \alpha^{(v)}_k,r\right\rangle \\
		=&
		\frac{1}{\mathcal{N}(u)\mathcal{N}(v)}\sum_{k,l=0}f^*\left(\alpha^{(u)}_k,r\right)f\left(\alpha^{(v)}_l,r\right)G_j\left(\beta_{k}^{(u)},\beta_{l}^{(v)}\right)\exp\left[-\frac{\left\vert\beta_{k}^{(u)}\right\vert^{2}+\left\vert\beta_{l}^{(v)}\right\vert^{2}}{2}+\beta_{k}^{(u)*}\beta_{l}^{(v)}\right].
	\end{split}
\label{kl_equation}
\end{equation}
The following is an analytic formulation of $ G_j\left(\beta_{k}^{(u)},\beta_{l}^{(v)}\right)$. We have assumed $\lambda=\cosh(r)$ and $\lambda_1=\sinh(r)$ for more concise expressions.
\\
1. For the operator $ \hat{K}^{\dagger}_i\hat{K}_j=\hat{I}$, we obtain
\begin{equation}
G_j\left(\beta_{k}^{(u)},\beta_{l}^{(v)}\right)=1;
\label{identity}
\end{equation}
2. For the operator $ \hat{K}^{\dagger}_i\hat{K}_j=\hat{a}$, we obtain 
\begin{equation}
G_j\left(\beta_{k}^{(u)},\beta_{l}^{(v)}\right)
=\lambda\beta_{l}^{(v)}-\lambda_1\beta_{k}^{(u)*};
\label{destroy}
\end{equation}
3. For the operator $ \hat{K}^{\dagger}_i\hat{K}_j=\hat{a}^{\dagger}\hat{a} $, we obtain 
\begin{equation}
	G_j\left(\beta_{k}^{(u)},\beta_{l}^{(v)}\right)=
\beta_{l}^{(v)}\beta_{k}^{(u)*} \left(\lambda^2+\lambda_1^2\right)+\lambda_1 \left[\lambda_1-\lambda\left(\beta_{l}^{(v)}\right)^2 \right]-\lambda\lambda_1\left(\beta_{k}^{(u)*}\right)^2; 
\label{photon_number}
\end{equation}
4. For the operator $ \hat{K}^{\dagger}_i\hat{K}_j=\hat{a}^{\dagger}\hat{a}^2 $, we obtain 
\begin{equation}
	\begin{split}
	G_j\left(\beta_{k}^{(u)},\beta_{l}^{(v)}\right)
		=&
		-\lambda_{1}\left(2\lambda^{2}+\lambda_{1}^{2}\right)\left(\beta_{k}^{(u)*}\right){}^{2}\beta_{l}^{(v)}+\left(\lambda^{2}+2\lambda_{1}^{2}\right)\beta_{k}^{(u)*}\left[\lambda\left(\beta^{(v)}_{l}\right)^{2}-\lambda_{1}\right]+\lambda\lambda_{1}^{2}\left(\beta_{k}^{(u)*}\right){}^{3}
		\\
		&+\lambda\lambda_{1}\beta_{l}^{(v)}\left[3\lambda_{1}-\lambda\left(\beta_{l}^{(v)}\right)^{2}\right];
	\end{split}		
\label{na}
\end{equation}
5. For the operator $ \hat{K}^{\dagger}_i\hat{K}_j=(\hat{a}^{\dagger}\hat{a})^2 $, we obtain  
\begin{equation}
	\begin{split}
	G_j\left(\beta_{k}^{(u)},\beta_{l}^{(v)}\right)
		=&-2 \lambda  \lambda _1 \left(\lambda ^2+\lambda _1^2\right) \left(\beta^{(u)*} _{k}\right)^3 \beta _{l}^{(v)}+\left(\beta^{(u)*} _{k}\right)^2 \left[\lambda ^4 \left(\beta _{l}^{(v)}\right)^2+4\lambda _1^2 \lambda ^2 \left(\beta _{l}^{(v)}\right)^2-2 \lambda _1 \lambda ^3-4 \lambda _1^3 \lambda
		\right.
		\\
		&\left.
		+\lambda _1^4 \left(\beta _{l}^{(v)}\right)^2 \right]+\beta^{(u)*}_{k} \beta _{l}^{(v)} \left[-2 \lambda _1 \lambda ^3 \left(\beta_{l}^{(v)}\right)^2-2 \lambda _1^3 \lambda \left(\beta _{l}^{(v)}\right)^2+\lambda ^4+8 \lambda _1^2 \lambda ^2+3 \lambda _1^4\right]
		\\
		&+\lambda ^2 \lambda _1^2 \left(\beta^{(u)*} _{k}\right)^4+\lambda _1 \left\lbrace-2 \lambda ^3 \left(\beta _{l}^{(v)}\right)^2+\lambda _1 \lambda ^2 \left[\left(\beta _{l}^{(v)}\right)^4+2\right]-4 \lambda _1^2 \lambda  \left(\beta _{l}^{(v)}\right)^2+\lambda _1^3\right\rbrace;
	\end{split}	
\label{nn}	
\end{equation}
6. For the operator $\hat{K}^{\dagger}_i\hat{K}_j=\left(\hat{a}^{\dagger}\hat{a}\right)^2\hat{a} $, we obtain 
\begin{equation}
	\begin{split}
	G_j\left(\beta_{k}^{(u)},\beta_{l}^{(v)}\right)
		=&\lambda  \lambda _1^2 \left(3 \lambda ^2+2 \lambda _1^2\right) \left(\beta^{(u)*}_{k}\right)^4 \beta _{l}^{(v)}-\lambda _1 \left(\beta^{(u)*}_{k}\right)^3 \left[3 \lambda ^4 \left(\beta _{l}^{(v)}\right)^2
		+6 \lambda _1^2 \lambda ^2 \left(\beta _{l}^{(v)}\right)^2-4 \lambda _1 \lambda ^3\right.
		\\
		&\left.+\lambda _1^4 \left(\beta _{l}^{(v)}\right)^2
		-6 \lambda _1^3 \lambda \right]+\left(\beta^{(u)*}_{k}\right)^2 \beta_{l}^{(v)} \left[\lambda^5 \left(\beta _{l}^{(v)}\right)^2+3\lambda _1^4 \lambda\left(\beta _{l}^{(v)}\right)^2-5 \lambda _1 \lambda^4-20 \lambda _1^3 \lambda ^2
		\right.
		\\
		&\left.
		+6 \lambda _1^2 \lambda ^3 \left(\beta _{l}^{(v)}\right)^2
		-5 \lambda _1^5  \right]+\beta^{(u)*}_{k} \left\lbrace\lambda ^5 \left(\beta _{l}^{(v)}\right)^2-\lambda _1 \lambda ^4 \left[2\left(\beta _{l}^{(v)}\right)^4+1\right]+16\lambda _1^2 \lambda ^3\left(\beta _{l}^{(v)}\right)^2\right.
		\\
		&\left.
		-\lambda _1^3 \lambda ^2 \left[3 \left(\beta _{l}^{(v)}\right)^4+10\right]+13 \lambda _1^4 \lambda  \left(\beta _{l}^{(v)}\right)^2-4 \lambda _1^5\right\rbrace-\lambda ^2 \lambda _1^3 \left(\beta^{(u)*}_{k}\right)^5
		\\
		&
		+\lambda  \lambda _1 \beta _{l}^{(v)} \left\lbrace-2 \lambda ^3 \left(\beta _{l}^{(v)}\right)^2+\lambda _1 \lambda ^2 \left[\left(\beta _{l}^{(v)}\right)^4+6\right]-8 \lambda _1^2 \lambda  \left(\beta _{l}^{(v)}\right)^2+9 \lambda _1^3\right\rbrace;
	\end{split}		
\label{nna}
\end{equation}
7. For the operator $ \hat{K}^{\dagger}_i\hat{K}_j=(\hat{a}^{\dagger}\hat{a})^3 $, we obtain 
	\begin{equation}
	\begin{split}
	G_j\left(\beta_{k}^{(u)},\beta_{l}^{(v)}\right)
		=&3 \lambda ^2 \lambda _1^2 \left(\lambda ^2+\lambda _1^2\right) \left(\beta^{(u)*}_{k}\right)^5 \beta _{l}^{(v)}-3 \lambda  \lambda _1 \left(\beta^{(u)*} _{k}\right)^4 \left[\lambda ^4 \left(\beta^{(v)}_{l}\right)^2+3 \lambda _1^2 \lambda ^2\left(\beta^{(v)} _{l}\right)^2-2 \lambda _1 \lambda ^3 
		\right.
		\\
		&\left.
		+\lambda _1^4 \left(\beta^{(v)}_{l}\right)^2-3 \lambda _1^3 \lambda \right]+\left(\beta^{(u)*} _{k}\right)^3 \beta^{(v)}_{l} \left[\lambda ^6 \left(\beta^{(v)}_{l}\right)^2+9 \lambda _1^2 \lambda ^4 \left(\beta^{(v)} _{l}\right)^2+9 \lambda _1^4 \lambda ^2 \left(\beta^{(v)}_{l}\right)^2
		\right.
		\\
		&\left.
		+\lambda_1^6\left(\beta^{(v)} _{l}\right)^2
		-9 \lambda _1 \lambda ^5-36 \lambda _1^3 \lambda ^3-15 \lambda _1^5 \lambda \right]+\left(\beta^{(u)*}_{k}\right)^2 \left\lbrace3 \lambda ^6 \left(\beta ^{(v)}_{l}\right)^2-\lambda _1 \lambda ^5 \left[3 \left(\beta^{(v)}_{l}\right)^4+4\right]
		\right.
		\\
		&\left.
		+6 \lambda _1^6 \left(\beta^{(v)} _{l}\right)^2
	+36 \lambda _1^2 \lambda ^4 \left(\beta^{(v)} _{l}\right)^2-\lambda _1^3 \lambda ^3 \left[9 \left(\beta^{(v)}_{l}\right)^4+28\right]-\lambda _1^5 \lambda  \left[3 \left(\beta^{(v)} _{l}\right)^4+13\right]
	\right.
	\\
	&\left.
	+45 \lambda _1^4 \lambda ^2 \left(\beta^{(v)}_{l}\right)^2\right\rbrace+\beta ^{(u)*}_{k} \beta^{(v)} _{l} \left\lbrace-9 \lambda _1 \lambda ^5 \left(\beta^{(v)} _{l}\right)^2				
	+\lambda _1^2 \lambda ^4 \left[3 \left(\beta^{(v)} _{l}\right)^4+32\right]\right.
		\\
		&\left.-36 \lambda _1^3 \lambda ^3 \left(\beta^{(v)}_{l}\right)^2+\lambda _1^4 \lambda ^2 \left[3 \left(\beta^{(v)} _{l}\right)^4+50\right]-15 \lambda _1^5 \lambda \left(\beta^{(v)} _{l}\right)^2+\lambda^6+7 \lambda_1^6\right\rbrace-\lambda ^3 \lambda _1^3 \left(\beta ^{(u)*}_{k}\right)^6
				\\
		&
		+\lambda _1 \left\lbrace-4 \lambda ^5 \left(\beta^{(v)}_{l}\right)^2
		+2 \lambda_1 \lambda ^4 \left[3 \left(\beta^{(v)} _{l}\right)^4+2\right]-\lambda _1^2 \lambda ^3 \left(\beta^{(v)}_{l}\right)^2 \left[\left(\beta ^{(v)}_{l}\right)^4+28\right]
			\right.
			\\
			&	\left.
		+\lambda _1^3 \lambda ^2 \left[9 \left(\beta^{(v)} _{l}\right)^4+10\right]-13 \lambda _1^4 \lambda \left(\beta^{(v)}_{l}\right)^2+\lambda _1^5\right\rbrace;
	\end{split}	
\label{nnn}	
\end{equation}
8. For the operator $ \hat{K}^{\dagger}_i\hat{K}_j=(\hat{a}^{\dagger}\hat{a})^4 $, we obtain 
	\begin{align}
		G_j\left(\beta_{k}^{(u)},\beta_{l}^{(v)}\right)
		=&\lambda ^4 \lambda _1^4 \left(\beta^{(u)*}_{k}\right)^8-4 \lambda ^3 \lambda _1^3 \left(\lambda ^2+\lambda _1^2\right) \beta^{(v)}_{l} \left(\beta^{(u)*}_{k}\right)^7+2 \lambda ^2 \lambda _1^2 \left[3 \left(\beta^{(v)}_{l}\right)^2 \lambda ^4+8 \lambda _1^2 \left(\beta^{(v)}_{l}\right)^2 \lambda ^2
		\right.\nonumber
		\\
		&\left.
		-6 \lambda _1 \lambda ^3-8 \lambda _1^3 \lambda +3 \lambda _1^4 \left(\beta^{(v)}_{l}\right)^2\right] \left(\beta^{(u)*}_{k}\right)^6-2 \lambda  \lambda _1 \beta^{(v)}_{l} \left[2 \left(\beta^{(v)}_{l}\right)^2 \lambda ^6+12 \lambda _1^2 \left(\beta^{(v)}_{l}\right)^2 \lambda ^4
\right.\nonumber \displaybreak[1]
		\\
		&\left.
		-15 \lambda_1 \lambda^5-48 \lambda _1^3 \lambda ^3+12 \lambda _1^4 \left(\beta^{(v)}_{l}\right)^2 \lambda^2-21 \lambda_1^5 \lambda +2 \lambda _1^6 \left(\beta^{(v)}_{l}\right)^2\right] \left(\beta^{(u)*}_{k}\right)^5+\left\lbrace\left(\beta^{(v)}_{l}\right)^4 \lambda ^8
		\right.\nonumber
		\\
		&\left.-192 \lambda _1^5 \left(\beta^{(v)}_{l}\right)^2 \lambda ^3 -24 \lambda _1 \left(\beta^{(v)}_{l}\right)^2 \lambda ^7
+4 \lambda _1^4 \left[9 \left(\beta^{(v)} _{l}\right)^4+31\right] \lambda ^4+\lambda _1^8 \left(\beta^{(v)}_{l}\right)^4
		\right.\nonumber
		\\
		&\left.+4 \lambda _1^2 \left[4 \left(\beta^{(v)} _{l}\right)^4+7\right] \lambda ^6
		-168 \lambda _1^3 \left(\beta^{(v)}_{l}\right)^2 \lambda ^5
		+2 \lambda _1^6 \left[8 \left(\beta^{(v)}_{l}\right)^4+29\right] \lambda ^2
			\right.\nonumber\displaybreak[1]
		\\
		&\left.	-36 \lambda _1^7 \left(\beta^{(v)}_{l}\right)^2\lambda			
		 \right\rbrace \left(\beta^{(u)*}_{k}\right)^4
		-2 \beta^{(v)}_{l} \left\lbrace2 \lambda _1 \left[\left(\beta^{(v)}_{l}\right)^4+8\right] \lambda ^7+2 \lambda _1^3 \left[6 \left(\beta^{(v)} _{l}\right)^4+79\right] \lambda ^5
		\right.\nonumber\displaybreak[1]
\\
&\left.-3 \left(\beta^{(v)} _{l}\right)^2 \lambda ^8-144 \lambda _1^4 \left(\beta^{(v)}_{l}\right)^2 \lambda ^4
+2 \lambda _1^5 \left[6 \left(\beta^{(v)} _{l}\right)^4+103\right] \lambda ^3+2 \lambda _1^7 \left[\left(\beta^{(v)}_{l}\right)^4+20\right] \lambda	\right.	\nonumber	\displaybreak[1]	 \\
				 &
				 \left.-56 \lambda _1^2 \left(\beta^{(v)} _{l}\right)^2 \lambda^6-72 \lambda _1^6 \left(\beta^{(v)}_{l}\right)^2 \lambda ^2-5 \lambda _1^8 \left(\beta^{(v)} _{l}\right)^2\right\rbrace \left(\beta^{(u)*}_{k}\right)^3
				 				 +\left\lbrace7 \left(\beta^{(v)}_{l}\right)^2 \lambda ^8
		\right.\nonumber\displaybreak[1]
		\\
		&\left.+2 \lambda _1^2 \left(\beta^{\left(v\right)}_{l}\right)^2 \left[3 \left(\beta^{(v)}_{l}\right)^4+106\right] \lambda ^6
		-8 \lambda _1 \left[3 \left(\beta^{\left(v\right)} _{l}\right)^4+1\right] \lambda ^7
		-24 \lambda _1^3 \left[7 \left(\beta^{(v)} _{l}\right)^4+6\right] \lambda ^5
		\right.
		\\
		&\left.+4 \lambda _1^4 \left(\beta^{(v)}_{l}\right)^2 \left[4 \left(\beta^{(v)}_{l}\right)^4+165\right]\lambda ^4
		+2 \lambda _1^6 \left(\beta^{(v)}_{l}\right)^2 \left[3 \left(\beta^{(v)} _{l}\right)^4+178\right] \lambda ^2+25 \lambda _1^8 \left(\beta^{(v)}_{l}\right)^2
	 \right.\nonumber
				\\
		&\left.	-4 \lambda _1^7 \left[9 \left(\beta^{(v)} _{l}\right)^4+10\right]\lambda-12 \lambda _1^5 \left[16 \left(\beta^{(v)} _{l}\right)^4+19\right] \lambda ^3
		\right\rbrace \left(\beta^{(u)*} _{k}\right)^2+ \left\lbrace
		-32 \lambda _1 \left(\beta^{(v)} _{l}\right)^2 \lambda ^7 
		\right.\nonumber\displaybreak[1]
		\\
		&\left.
		+6 \lambda _1^2 \left[5 \left(\beta^{(v)} _{l}\right)^4+18\right] \lambda ^6 
		+24 \lambda _1^4 \left[4 \left(\beta^{(v)} _{l}\right)^4+19\right] \lambda ^4
		-4 \lambda _1^5 \left(\beta^{(v)}_{l}\right)^2 \left[\left(\beta^{(v)}_{l}\right)^4+103\right] \lambda ^3
		\right.\nonumber\displaybreak[1]
		\\
		&
		\left.+\lambda ^8 +15 \lambda _1^8+2 \lambda _1^6 \left[21 \left(\beta^{(v)} _{l}\right)^4+130\right] \lambda ^2-4 \lambda _1^3 \left(\beta^{(v)} _{l}\right)^2 \left[\left(\beta^{(v)}_{l}\right)^4+79\right] \lambda ^5
		\right.\nonumber\displaybreak[1]
		\\ 
		&		
		\left.   -80 \lambda _1^7 \left(\beta^{(v)}_{l}\right)^2 \lambda\right\rbrace\beta^{(v)}_{l} \beta^{(u)*}_{k}+\lambda _1 \left\lbrace
		-12 \lambda _1^2 \left(\beta^{(v)}_{l}\right)^2 \left[\left(\beta^{(v)} _{l}\right)^4+12\right]\lambda ^5-40 \lambda _1^6 \left(\beta^{(v)}_{l}\right)^2 \lambda
			\right.\nonumber\displaybreak[1]
		\\
		&\left. -8 \left(\beta^{(v)}_{l}\right)^2 \lambda ^7+\lambda _1^3 \left[\left(\beta^{(v)}_{l}\right)^8+124\left(\beta^{(v)}_{l}\right)^4+60\right] \lambda ^4
		-4 \lambda _1^4 \left(\beta^{(v)}_{l}\right)^2 \left[4 \left(\beta^{(v)}_{l}\right)^4+57\right] \lambda ^3
			\right.\nonumber
		\\ 
		&\left.
		+4 \lambda _1 \left[7\left(\beta^{(v)} _{l}\right)^4+2\right]\lambda ^6+2 \lambda _1^5 \left[29\left(\beta^{(v)} _{l}\right)^4+18\right] \lambda ^2+\lambda _1^7\right\rbrace.\nonumber
\label{nnnn}		
\end{align}

\section{The optimal recovery channel}
\label{optimal_recover_channel}
Here we elaborate the numerical method to identify the optimal recovery channel. We assume that the error space consists of the bases {$\vert i^{(u)}\rangle=A_i\vert u_L\rangle/\sqrt{\langle u_L \vert A^{\dagger}_iA_i\vert u_L\rangle}$, where \(\hat{A}_i\) are the Kraus operators of the error channel and \(\vert u_L\rangle\) is the logical basis}. Therefore, the recovery operators are  linear superpositions of the operators $\hat{B}_i$,
\begin{equation}\label{key2}
	\hat{R}_k=\sum_i x_{k,i} \hat{B}_i, 	
\end{equation}
where we have defined the operator $\hat{B}_{i}\in\lbrace \vert 0_{\text{L}}\rangle\langle i^{(u)}\vert,~\vert 1_{\text{L}}\rangle\langle i^{(u)}\vert\rbrace$ to restore the encoded information from the error space into the logical space. We can find the optimal coefficients $x_{k,i}$ to obtain the optimal recover operators $\hat{R}_{k}$. To this end, we need to maximize the entanglement fidelity by optimizing the coefficient $x_{k,i}$~\cite{PhysRevLett.94.080501,PhysRevA.65.030302,PhysRevA.75.012338,Kosut2009Oct},
\begin{equation}\label{channel_fidelity}
	F= \frac{1}{4}\sum_{ij}\left\vert\Tr\left\lbrace \hat{R}_i\hat{A}_j\right\rbrace \right\vert^2, 	
\end{equation}
which is equivalent to solving the following convex semidefinite program
\begin{equation}\label{key}
	\begin{split}
		&X_{\text{opt}}=\underset{X}{\operatorname{argmax}}~\frac{1}{4}\Tr\left\lbrace XW\right\rbrace, \,\,\,\,\,\,\, \text{with} \,\,\,\,\,\,\, \sum_{ij}X_{ij}\hat{B}^{\dagger}_i\hat{B}_j=\hat{I},~~~~X\succcurlyeq 0,	
	\end{split}
\end{equation}
where the elements $ X_{ij} $ and $W_{ij}$ of the matrix  $ X $ and $W$ can be written as 
\begin{equation}
	X_{ij}=\sum_{l}x_{ri}x^*_{rj}, \,\,\,\,\,\,\, W_{ij}=\Tr\left[\mathcal{N}\left(\hat{B}_i\otimes \hat{B}^{\dagger}_j\right)\right], 
\end{equation}
respectively. From the solution to the above  convex optimization problem we acquire the corresponding optimal recovery channel, where the optimal recovery operators can be given by a singular value decomposition of $X_{\text{opt}}$,
\begin{equation}
	\begin{split}
		X_{\text{opt}}&=V\Omega V^{\dagger}, \\
		R_{\text{opt}}&=\sqrt{\Omega_r}\sum_iV_{ir}B_i,		
\end{split}
\end{equation}
where $\Omega_r$ is the singular value and $V$ is an unitary matrix. We utilize Cvxpy for semidefinite convex optimization in Python~\cite{diamond2016cvxpy,agrawal2018rewriting}. {This recovery channel represents the optimal recovery channel, which defines the upper bound for QEC. As an example, we calculate the  upper fidelity bounds for the optimal and conventional GKP codes at the time scales $\kappa \tau = 0.0004$ and $\kappa_{\phi} \tau = 0.0004/1.5$, as shown in Tab.~\ref{table}. Our results show that, although both the optimal and conventional codes exhibit some noise bias for different encoded states, due to the incomplete satisfaction of the KL condition, the optimal code surpasses the conventional code for all six states. 
}
\begin{table*}[htbp]
	\centering
	\caption{Comparison of the fidelity upper bounds for the optimal and conventional GKP codes}
	\renewcommand{\arraystretch}{1.5}
	\begin{ruledtabular}
			\begin{tabular}{r|@{}c*{5}{@{\hspace{0.3pt}}c}}
				Encoding state & $\vert 0_{\text{L}}\rangle$  & $\vert 1_{\text{L}}\rangle$  & $\left(\vert 0_{\text{L}}\right\rangle+\vert 1_{\text{L}}\rangle)/\sqrt{2}$ & $\left(\vert 0_{\text{L}}\right\rangle-\vert 1_{\text{L}}\rangle)/\sqrt{2}$& $\left(\vert 0_{\text{L}}\right\rangle+i\vert 1_{\text{L}}\rangle)/\sqrt{2}$& $\left(\vert 0_{\text{L}}\right\rangle-i\vert 1_{\text{L}}\rangle)/\sqrt{2}$\\
				\cline{1-7}
				Optimal GKP &0.999968 & 0.999978 & 0.999964&0.999957 &0.999954 & 0.999954   \\
				Conventional GKP&0.999918 & 0.999926 & 0.9999251& 0.999919& 0.999919& 0.999919
			\end{tabular}
	\end{ruledtabular}
	\label{table}
	\vspace{-1.8\baselineskip}
\end{table*}

\begin{figure}[tpb]
	\centering
	\includegraphics[width=1\columnwidth]{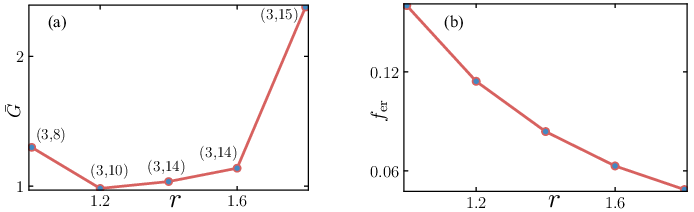}
	\caption{(a) The gain of optimal GKP codewords relative to the best conventional GKP codewords versus the squeezing amplitude $r$. Here, $(M_{\text{o}},~M_\text{c})$ represents the fewest $M$ values needed by the optimized GKP and the best conventional GKP codewords, respectively. The gains are above one for various squeezing amplitudes $r$, showing that the optimal GKP codewords significantly reduce the necessary $M$ while maintaining higher error-correctability. (b) The distance between the stabilizer operators of the ideal and optimal approximate GKP codewords versus the squeezing strength. {We set a target limit of $L_{\text{st}} \sim 10^{-3}$.}}. 
	\label{fig1_noize}
\end{figure}
\begin{figure}[tpb]
	\centering
	\includegraphics[width=1\columnwidth]{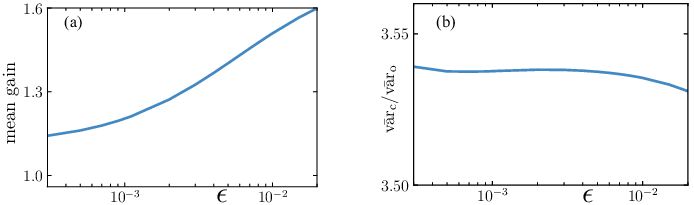}
	\caption{(a) Mean gain of optimal GKP codewords versus conventional GKP codewords as a function of the noise magnitude of the coefficients $c^{(u)}_k$. (b) Mean-variance ratio of optimal and conventional approximate GKP codewords versus noise size $\epsilon$.} 
	\label{fig1_ni}
\end{figure}

\section{Universality of the neural-network-based GKP codewords}

Here, we investigate the universality of the neural network-based GKP codewords, including the performance of these approximate GKP codewords under various squeezing strengths, the impact of small perturbations in the codeword coefficients, and the performance when the codeword coefficients are restricted to real numbers.
\subsection{Universality for various squeezing strengths}
To illustrate the university of our proposal, we investigate the minimum number of squeezed coherent states required to optimize GKP under different squeezing levels. Our proposal effectively reduces the number of squeezed coherent states and enhances error correction performance under different strength squeezing levels, as shown in Fig.~\ref{fig1_noize}(a). Specifically, the optimized approximate GKP codewords lower the number of squeezed states to \textit{seven} in each logical base ($1/3$ of the conventional approximate GKP for squeezing amplitudes $r>1$) while maintaining better error correction performance. 

{The ideal stabilizers only offer rough stabilization for the approximate code at finite squeezing degrees. To address this, we propose more precise stabilizer operators, inspired by the conventional GKP code in the limit $\zeta^{-1} \gg 1$,}
\begin{equation}
	\begin{split}
		\hat{S}_{q,\text{ap}}=\exp\left[i2\sqrt{\pi}\left(f_{11}\hat{q}+f_{12}\hat{p}\right)\right],\\ \hat{S}_{p,\text{ap}}=\exp\left[-i2\sqrt{\pi}\left(f_{21}\hat{q}+f_{22}\hat{p}\right)\right],
	\end{split}
	\label{stab}
\end{equation}
where the elements of the coefficient matrix $\textbf{f}=[f_{11},~f_{12};~f_{21},~f_{22}]$ can be complex numbers.
The matrix elements have the relation $f_{11}f_{22}-f_{12}f_{21}=1$, which preserves the commutation $\hat{S}_{q,\text{ap}}\hat{S}_{p,\text{ap}}=\hat{S}_{p,\text{ap}}\hat{S}_{q,\text{ap}}$. 
 {
 	
We assume that the optimized stabilizers also follow this general form, with $\vert u_L \rangle = \hat{S}_{p,\text{ap}} \vert u_L \rangle$ and $\vert u_L \rangle = \hat{S}_{q,\text{ap}} \vert u_L \rangle$. Satisfying this condition is equivalent to satisfying the relations $\langle u_L \vert \hat{S}_{q,\text{ap}} \vert u_L \rangle \approx 1$, $\langle u_L \vert \hat{S}_{p,\text{ap}} \vert u_L \rangle \approx 1$, $\langle u_L \vert \hat{S}^{\dagger}_{q,\text{ap}} \hat{S}_{q,\text{ap}} \vert u_L \rangle \approx 1$, and $\langle u_L \vert \hat{S}^{\dagger}_{p,\text{ap}} \hat{S}_{p,\text{ap}} \vert u_L \rangle \approx 1$. To quantify how well the codewords satisfy these conditions, we define the following cost function, 	
\begin{equation}
	\begin{split}
		L_{\text{st}}=\sum_{u=0,1}\sum_{\hat{O}}\vert 1-\langle u_L\vert\hat{O}\vert u_L\rangle\vert^2,
	\end{split}
\end{equation}
where $\hat{O}\in \left\lbrace\hat{S}_{q,\text{ap}},~\hat{S}_{p,\text{ap}},~\hat{S}^{\dagger}_{q,\text{ap}}
\hat{S}_{q,\text{ap}},~ \hat{S}^{\dagger}_{p,\text{ap}}\hat{S}_{p,\text{ap}}\right\rbrace$. Therefore, we incorporate the above cost function into the total loss function. It is important to note that the Pauli operators in the logical code space are represented as $\hat{\sigma}_z = \hat{S}^{1/2}_{q,\text{ap}}$ and $\hat{\sigma}_x = \hat{S}^{1/2}_{p,\text{ap}}$, which can converge autonomously without the need to consider an additional loss function.

}

The coefficient matrix $f$ for ideal GKP codewords has the elements $f_{\text{11}}=f_{\text{22}}=1$ and $f_{\text{12}}=f_{\text{21}}=0$. We define $f_{\text{er}}$ as the distance between the stabilizer operators of the approximate and the ideal GKP codewords
\begin{equation}
	f_{\text{er}}=\left\vert f_{11}-1\right\vert+\left\vert f_{22}-1\right\vert+\left\vert f_{12}\right\vert+\left\vert f_{21}\right\vert.
\end{equation}
In Fig.~\ref{fig1_noize}(b), we simulate $f_{\text{er}}$ as a function of the squeezing strength. As the squeezing increases, $f_{\text{er}}$ decreases monotonically, showing that these stabilizer operators become closer to the stabilizer operators of the ideal codewords.
Therefore, the optimal GKP codewords are valid and useful at different squeezing levels.

\subsection{Robustness under disturbances}
We investigate how profile perturbations of the coefficient $c^{(u)}_k$ impact the error correction performance of the encoded quantum states. The coefficients of the logical basis can be expressed as the mean values with small fluctuations due to imperfect control
\begin{equation}
	c^{(u)}=c^{(u)}+\epsilon c^{(u)}\varXi,
\end{equation}
where $\epsilon$ is the magnitude of the noise with a value in { $\epsilon\in(0,0.02]$}, and {$\varXi$ is a random matrix whose elements are sampled from a uniform distribution over the interval $\Xi\in[-0.5,0.5]$}.
Since the mean infidelity is proportional to $\bar{L}_{\text{er}}$, we can define the average gain
\begin{equation}
 \bar{G}=\bar{L}_{\text{er}}(\text{ogkp})/\bar{L}_{\text{er}}(\text{cgkp}),
 \end{equation}
  of the optimal GKP codewords compared to the best conventional ones.  We show the average gain and the ratio $\bar{\text{var}}_c(\bar{L}_{\text{er}})/\bar{\text{var}}_o(\bar{L}_{\text{er}})$ for randomly generated noise $\varXi$ versus the noise magnitude $\epsilon$ in the Fig.~\ref{fig1_ni} (a) and (b). The mean gains are always above one and increase with increasing noise magnitude $\epsilon$; the variance of the loss function for the conventional GKP codewords remains more than 3.5 times that of the optimized GKP one. {Our encoding demonstrates significantly greater robustness, whereas the conventional GKP codes are highly susceptible to noise, with the imperfect state preparation severely impairing error correction. This advantage arises from our codes avoiding 2/3 of the large-amplitude squeezed coherent states that are critical for the conventional codes, but challenging to prepare accurately. These states, particularly in optical systems, are regarded as a fundamental obstacle to the GKP state preparation. Thus, our codes enhance the stability and simplifies the state preparation, substantially reducing the impact of imperfections.}
  This demonstrate that the noise resilience of the \textit{optimized} GKP encoding outperforms that of the \textit{conventional} GKP encoding, allowing for effective error correction and reducing the demand for extremely precise control to generate the GKP codewords.

\subsection{The real coefficient GKP codewords}
Here we restrict to the real coefficients $c^{(u)}_k$ to investigate the neural-network based GKP codewords. Our results show that the optimal GKP codewords at squeezing level $r=1.1$ still exhibit better quantum error correction ability than the best conventional one, 
where the number of squeezing states is $7$ and  corresponding coefficients are 
\begin{figure}[htbp]
	\centering
	\includegraphics[width=1\columnwidth]{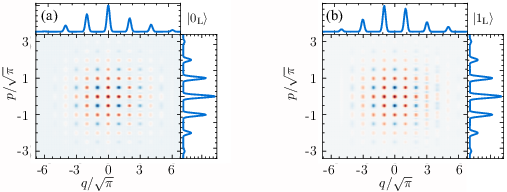}
	\caption{Panels (a) and (b) give the Wigner functions for the optimal codewords $\vert 0_L\rangle$ and $\vert1_L\rangle$ with real coefficients $c^{(u)}_k$ for $M=3$, respectively.} 
	\label{fig1_wigner}
\end{figure}
 {
\begin{equation}
	\begin{split}
		&c^{(0)}/\mathcal{N}(0)=[0.054826
		,~ 0.228328       
		,~ 0.381576       
		,~ 0.470909       
		,~ 0.334463       
		, ~0.243658       
		, ~0.118351], \\
		&c^{(1)}/\mathcal{N}(1)=[ 0.114688,  ~
		 0.258726,  
		 ~0.375942,  
		 ~0.354715,  
		 ~0.229235, 
		~ 0.163002,  
		~-0.039539].
	\end{split}
\label{real_coe}
\end{equation}}
For these optimal codewords, the stabilizer operators match those in Eq.~(\ref{stab}), { with the coefficient matrix \begin{equation} f = \left[ \begin{matrix}0.999531+0.000695i & -0.000088 + 0.110767i \\ -0.000067 -0.032467i & 1.004067 - 0.000703i \end{matrix} \right]\end{equation} having an approximation level close to \(L_{\text{st}}\sim 10^{-3}\). }
Note that the error correction performance of this GKP state is only slightly poorer than that of the GKP state with complex coefficients offered in the main text. Moreover, we show the Wigner function of codewords with the real coefficients Eq.~(\ref{real_coe}) in Figs.~\ref{fig1_wigner} (a) and (b). Clearly, the projection of the Wigner function in the momentum coordinate system is consistent with that of the complex field. The main difference is the projection distribution in the position coordinates. This shows that optimizing GKP codewords for real coefficients is also a reliable and near-optimal alternative.


%

